\documentclass[12pt]{article}
\usepackage{latexsym}
\renewcommand{\epsilon}{\varepsilon} 
\begin{document} 
\title{Why Quantum Information Processing} 
\author{Hoi-Kwong Lo\footnote{email: hklo@comm.utoronto.ca} \\ 
Department of Electrical and Computer
Engineering; \\
and Department of Physics \\  University of Toronto \\ 10 King's College
Road, Toronto, ON Canada M5S 3G4} 
\date{\today} 
\maketitle 
\begin{abstract} 
 
In this brief note, I will consider the following questions:
(1)~ What is QIP? (2)~Why QIP is interesting? (3)~What QIP can do?
(4)~What QIP cannot do?
(5)~What are the major challenges in QIP?

\end{abstract} 
 
 {\par\medskip\noindent
\begin{minipage}[t]{5in} 
\end{minipage} 
 } 
 

\section{Organization of this course}

Chapter~1: This Introduction: Logistics, Motivation and Introduction.

Chapter~2: Basics of quantum mechanics and quantum information.

Chapter~3: Elements of quantum computing and quantum error correction

Chapter~4: Quantum source coding

Chapter~5: Quantum channel coding

Chapter~6: Quantum cryptography~I: Quantum key distribution

Chapter~7: Quantum cryptography~II: Bit commitment, secret sharing, etc.

Some guest lecturers may also be invited for special topics on
the subject.

\section{Introduction} 
\label{S:Intro} 

[N.B. This brief note summarizes my personal perspective on the subject.
Comments and criticisms are most welcome.]

Quantum information processing (QIP) is a new and exciting
area of inter-disciplinary research. It combines ideas in
physics, mathematics, computer science and engineering.
The main goal of QIP is to harness the fundamental laws of
quantum mechanics to improve dramatically all aspects (e.g. acquisition,
transmission, and processing) of information processing.
In this introductory chapter, I will survey the motivation and power of
QIP. More specifically, I will answer the following questions:

(1) What is QIP?

(2) Why QIP is interesting?

(3) What QIP can do?

(4) What QIP cannot do?

(5) What are the major challenges in QIP?

\subsection{What is QIP?}

Let me begin with the first question: What is QIP?
One can think of QIP as a new quantization program.
Its goal is combine and unify quantum mechanics with other areas of
research such as theory of computing, cryptography, information theory
and error-correcting codes. In the synthesis of the ideas of
quantum mechanics with other subjects, new research problems
appear. As a result of tackling those problems,
new insights are gained and a more complete
and unified theory emerge.

In some cases, such a synthesis can lead to the questioning of
the very foundation of the original subject.
Take the example of the theory of computation.
It was commonly accepted for a long time that the
Church-Turing thesis---that any computational model
can be {\it efficiently} simulated by a conceptual device
called the Turing machine---is a valid assumption/foundation of the
subject. The upshot is that it does not
really matter how a classical computer operates:
Classical computers of different architectural designs
essentially give equivalent computational power.
However, quantum computing challenges the Church-Turing
thesis: There is no known efficient classical algorithm for
simulating the evolution of a quantum system. Moreover,
in 1994, Peter Shor of AT\&T made a remarkable discovery \cite{shor} that
quantum computers can, in principle, factor large integers
efficiently (i.e., in polynominal time in the size of the inputs).
In contrast, no efficient classical algorithm for factoring is known.

Therefore, we have to ask the question:
Does the Church-Turing thesis apply to a quantum computer?
This is an important question. If we can answer this question
one way or another, we are making an important conceptual breakthrough.
Why is this so? Suppose the answer is yes. Then, there must
exist an efficient classical algorithm for simulating any
quantum system. Such a discovery will be a big surprise to
quantum physicists and will make physicists and computational
chemists who are interested in simulating quantum systems very
happy. I have been told that much of the world's computational
power is used in computational chemistry. And an important part of
computational chemistry is quantum chemistry. Added to this,
our colleagues in lattice QCD or condensed matter physics or
nanotechnologists will
be delighted to learn about such classical algorithms!
On the other hand, if the answer is no, then there will be a
burning desire for us to work on the construction of a quantum
computer, our ultimate computational machine.

Moreover, I remark that Shor's discovery is far more than an intellectual
curiosity. Indeed, the presumed hardness of factoring is
the foundation of security of RSA encryption
scheme, which is well-known standard encryption scheme.
Since RSA is widely used to guarantee the security in
electronic commerce and business, if a quantum computer is
ever built, the whole foundation of security of electronic
commerce and business will fall apart.

In the context of information theory, our viewpoint is that
there is only one information theory---quantum information theory.
Conventional information
theory is simply a special case of quantum information theory.
We hope that quantum information theory will make
conventional information theory complete
in an analogous way that the theory of complex numbers makes real numbers
complete.

\subsection{Why QIP is interesting?}

\subsubsection{1.~Applications of the fundamental laws of nature
have been the driving force of economic growth.}
I am impressed by the viewpoint of Prof. Michael Berry \cite{berry}. 
Throughout history, life has been transformed by the conscious
application of physical laws. In the nineteenth century,
life was transformed by the Newton's laws and later
thermodynamics through heat engines, which were
the driving force of the whole industrial revolution.
Incidentally, Newton's laws and thermodynamics are
the foundations of mechanical engineering.
In the twentieth century, life was similarly transformed by
the application of Maxwell's equations, through electricity
and communicating words and pictures at the speed of
light through electromagnetic waves.
Incidentally, Maxwell's equations can be regarded
as an important foundation of electrical engineering.
Looking forward to the twenty-first century,
Michael Berry confidently proclaimed that: "It is easy to
predict that in the twenty-first century, it will be quantum
mechanics that influences all our lives."

He went to write:
"There is a sense in which quantum mechanics is already having profound
effects. Leon Lederman claims that a large part of the gross
national product of the industrial countries stems from quantum
mechanics. I suppose he is refering to transistors---the
`fundamental particles' of modern electronics---that depend on
properties of semiconducting materials designed by applying
quantum mechanics to electrons in solids, and to lasers, where the
Bose-Einstein statistics of identical particles generate coherent
avalanches of photons to read the bar-codes in our supermarkets and guide
delicate surgery in our eyes."

\subsubsection{2.~Predicted Demise of Moore's Law}
A main motivation of quantum information processing
is the predicted demise of
the Moore's law \cite{birnbaum}. Gordon Moore---Co-founder of Intel---made an
empirical observation that the number of transistors in a single
chip has been growing exponentially with time. (It doubles
every 18 to 24 months.) As a consequence, the computing
power in a small computer, such as our PC, also grows exponentially
with time. This exponentially grow is tied to the exponential
grow in the chip market, which has been doubling about every
five years.

Moore's law cannot continue forever.
There are two factors that will bring Moore's law to an end.
The first factor is economic. The cost of a fabrication facility
has also been increasing exponentially with time, doubling
every chip generation. This is called Moore's second law.
Currently, a single fabrication facility will take billions of
dollars to build. By 2010, according to Moore's second law,
it may cost about \$40 billion, which
is roughly 10 percent of the total annual market at that time.

The second factor is technological. Incidentally, the
Semiconductor Industry Association has published
a National Technological Roadmap that studies the continuation of
the current exponential growth in computational power until the year 2012.
The number of technological obstacles for achieveing such exponential
growth has been growing. Some of those obstacles are
commonly believed to of fundamental, rather than technical
origin. Indeed,
one can consider the microscopic implication of Moore's law.
If one plots the number of impurities in a single transistor,
one sees an exponential decrease with time. By extrapolation,
one sees that by the year 2020 or so, the number of impurities
per transistor is about one. That would mean that we
will be effectively working with a single electron transistor.
The operation of such a single-impurity/electron
transistor must be described directly by quantum mechanics.
Moreover, since electrons are indivisible, naive
extrapolation of (the microsopic implication of)
the Moore's law will make no sense at all beyond the year 2020.
The relevation is that quantum effects are going to play a
dominant role in the operations of future transistors.
A fundamental understanding of quantum effects will, therefore,
be necessary for the continued increase in computing power.

Information revolution of the twentieth century was based
on bulk quantum effects.
There has been growing appreciation that
this might only be the first phase of the information revolution.
In the second phase of the information revolution, which will take
place in the twenty-first century, it has been suggested that
quantum physics may play a direct role in the operation of
computing devices.

\subsubsection{3.~QIP offers
dramatic improvement over conventional information processing}

Quantum information processing is interesting because it
offers dramatic improvement over conventional information processing.
See Section~3.

\subsubsection{4.~QIP as a proof technique}
One argument against QIP is that large-scale quantum computation
may prove to be impractical. Moreover, even quantum mechanice
might turn out only to be an approximation of nature.
However, even in those cases,
QIP would remain an important proof techique
in solving mathematical problems.
For instance, in \cite{wolf}, QIP is used a proof technique to
obtain a better bound on conventional coding theory problem.
Indeed, this supports the viewpoint is one should consider
quantum information theory is a natural generalization of
classical information theory and as such it provides us with
new proof techniques in problems strictly in conventional information theory.

An analogy of quantum information
processing is complex numbers. Even though all measured quantities
are real (and we have yet to find a measurement that will
give directly a complex output!), the usefulness of complex analysis
is never in doubt!

\subsubsection{5.~QIP is exciting}
The final reason why QIP is interesting is that people often
find QIP
counter-intuitive, fascinating, provocative, and, most importantly,
fun to work on. It is a young and rapidly evolving subject which
may have lots of surprises waiting to be discovered.

Quantum mechanics is intellectually
interesting because it challenges our
way of thinking about the world. It changes the
rules of what is possible or what is
impossible. For instance,
in conventional information theory, we accept without question
that information can be copied and observed without change.
However, in quantum mechanics, there is a well-known principle
called the Heisenberg's uncertainty principle which asserts
that it is fundamentally impossible to know two conjugate
observables---e.g. its position and momentum---simultaneously.
As we will discuss later, an implication of the
Heisenberg's uncertainty principle is that quantum information
{\it cannot} be copied. This is contrary to our classical
intuition and is, in fact, the foundation of security of a new type of
cryptographic schemes based on quantum mechanics---quantum cryptography.
Heisenberg's uncertainty principle ensures that
an eavesdropper cannot copy quantum signals with
disturbing them. In other words, an eavesdropper will necessarily
leave her traces behind.

Quantum mechanics is one of the best battle-tested theory of the
physical world. It has routinely been used to describe microscopic
phenomena of nature inside atoms, lasers and solids.
So far, it has agreed prefectly with experiments.
Of course, every physical law is only a model of nature.
When extrapolated to domains far beyond its original conception,
it may eventually break down. Here, we are not concerned
with this potential breakdown. We will regard quantum
mechanics as a mathematical model and see how it influences
our thinking of information theory.

\subsection{What QIP can do?}
Quantum computing can efficiently solve some hard problems that
are intractable with conventional computers. A standard example is
Shor's efficient quantum algorithm for factoring,
which is a hard problem studied by mathematicians for centuries
and is also the foundation of security of standard RSA crypto-system.
Therefore, ``if a quantum computer is ever built, much of conventional
cryptography will fall apart!" (Brassard).

Fortunately, quantum mechanics can be used to make
codes as well as breaking them. Indeed, quantum cryptography
(the art of code-making), in principle, allows perfectly secure communication
between two parties in the presence of a technologically arbitrarily advanced
eavesdropper. Quantum code-making will be
discussed in Chapter~6.
In summary, quantum mechanics has the potential to revolutionize
both code-making and code-breaking.

Quantum information processing also allows other novel types of
information processing. For instance, in a process
called "quantum teleportation",
a quantum state is decomposed locally in one spatial location
and remotely reconstructed in another location through the
communication of a classicacl message and the prior sharing
of some quantum resource called "entanglement".
While the communication of the classical message ensures that
the process does not violate casuality, the very fact that the
state of an object can be decomposed in one place and re-constructed
in another even without knowing its complete description is a rather
surprising phenomenon. It reminds people of science
fiction and captures people's imagination in the
same way as teleportation in the TV series {\it Star Trek}.

Also, there are various multi-party computational or cryptographic
tasks that are impossible in the conventional setting. Yet, by allowing
the parties to share prior quantum correlations, they become
possible.

\subsection{What QIP cannot do?}

There are tasks that even quantum information processing cannot do.
The first example is to compute a ``non-computable function'' defined
in the standard Church-Turing model. So, while quantum computing
can speed up (sometimes dramatically) some computations, it
can always be simulated by a classical computer---but perhaps with an
exponential overhead in some resource---time or space or resolution.
 
A second example is a cryptographic task called quantum bit commitment.
This will be discussed in Chapter~7.

\subsection{What are the major challenges of QIP?}

A first major challenge of QIP is to discover new
tasks where QIP gives a dramatic advantage.
For instance, invent a useful new quantum algorithm.
One important open problem is the
graph isomorphism problem. Quantum algorithms will be
briefly discussed in Chapter~3.

A second major challenge of QIP is to construct a
large-scale quantum information processing in the real world.
While writing equations on paper is easy, building a real
large-scale quantum information processing is a major technological
challenge that many groups in the world have been working very hard on.
See Chapter~3 for a brief discussion and \cite{book} for details.

A third major challenge of QIP is to solve some of the
well-known open problems in the subject.
An example is to find an efficient way to
compute the various quantum channel capacities defined
in the literature. See Chapter~4.

A fourth major challenge of QIP is to continue the quantization
program and apply it to other subjects. If one takes the
quantization program seriously, one would believe that
it is meaningful to pick almost any scientific subject and ask if and
how it can be combined with quantum mechanics. For instance,
one can start with control theory and ask how quantum control theory
can be formulated.
[A caveat: It has been suggested that some subjects such as
game theory may not have a particularly useful or natural
quantum version. So, there may be limits to the viewpoint that
every subject can be quantized.]

A fifth major challenge is to use QIP as a proof technique to
tackle problems strictly in classical information theory.
The power and limitation of such a proof technique deserves
investigations.

\end{document}